\documentclass[aps,prb,twocolumn,showpacs,superscriptaddress]{revtex4}
\usepackage{bm}
\usepackage{mathptmx}
\usepackage{graphicx}

\newcommand{\smeq}{\! = \!}

\newcommand{\smmi}{\! - \!}

\newcommand{\be}{\begin{equation}}
\newcommand{\ee}{\end{equation}}
\newcommand{\bea}{\begin{eqnarray}}
\newcommand{\eea}{\end{eqnarray}}

\newcommand{\ci}{\mathrm{i}}
\begin{document}

\title{Slave boson theory for transport through magnetic molecules with vibronic states}
\author{M. D. Nu\~nez Regueiro}
\affiliation{Laboratoire de Physique des Solides, Universite Paris-Sud, Batiment 510,  91405-Orsay, France.}
\author{P. S. Cornaglia}
\affiliation{Centre de Physique Th\'eorique, \'Ecole Polytechnique, CNRS, 91128, Palaiseau Cedex, France.}
\author{Gonzalo Usaj}
\affiliation{Centro At\'omico Bariloche and Instituto Balseiro, 8400 San Carlos de Bariloche, Argentina.}
\author{C. A. Balseiro}
\affiliation{Centro At\'omico Bariloche and Instituto Balseiro, 8400 San Carlos de Bariloche, Argentina.}

\begin{abstract}
We study the electron transport through a magnetic molecular 
transistor in the Kondo limit using the slave boson technique. We include the electron-phonon coupling and analyze the cases where the spin of the molecule is either $S=1/2$ or $S=1$. We use the Schrieffer-Wolff transformation to write down a low energy Hamiltonian for the system. In the presence of electron-phonon coupling, and for $S\smeq1$, the resulting Kondo Hamiltonian has two active channels. At low temperature, these two channels interfere destructively, leading to a zero conductance. 
\end{abstract} 

\pacs{72.15.Qm, 73.22.-f}

\maketitle

\section{Introduction}

The technological advances of the last decades triggered a systematic study of 
electronic transport in nanoscale systems weakly connected to external
electrodes. Confinement of electrons at the nanoscale leads naturally to
energy level quantization and charging effects. In small quantum dots Coulomb blockade becomes a dominant effect as
shown in conductance experiments.\cite{Kastner92,AleinerBG02} This effect, together with the
enhancement of the low-temperature conductance in the valley between Coulomb
blockade peaks---a signature of the well known Kondo effect\cite{Glazman1988,Ng1988}---makes evident
that electron-electron correlations play a central role in these systems. 
Since the early experiments in quantum dots,\cite{Gordon1998,Wiel2000} the Kondo effect has also
been observed in a variety of molecules. The
conductance of molecular systems like C$_{60}$ molecules, Co and Cu complexes connected to metallic leads or carbon nanotubes, shows the signature of
Kondo-like behavior.\cite{WLiang_2002,JPark_2002,Yu2004,Yu2005} 
In these molecular junctions, the molecule is coupled
to the source and drain leads providing a path for charge transport. In some
cases a third gate electrode is used to create an external potential that
modifies the energy of the molecular orbitals. The use of normal,
ferromagnetic and superconducting electrodes gives rise to a rich behavior
and opens new alternatives for the study of the interplay between Kondo
screening and electronic correlations in the leads. In addition, transport through
highly structured molecules like the Mn$_{12}$, a single molecular magnet, has
been reported\cite{HeerscheGFZRWZBTC06} and the Kondo effect has been predicted to occur in this type of systems for both half-integer\cite{romeike:196601} and integer values of the molecular spin.\cite{LeuenbergerM06}

Despite of the substantial theoretical and experimental activity devoted to
the study of molecular transistors there are still many open questions. While
the theoretical frame for the study of the Kondo effect in semiconducting
quantum dots and molecular systems is basically the same, there are some
fundamental differences between them. Namely, vibrational modes can play a central role in the single electron transfer through molecules. In fact, molecules distort upon the addition or the
removal of electrons leading to a large electron-phonon interaction. As the
Coulomb charging energies in these systems can be considerably reduced by
screening due to the electrodes,\cite{Kubatkin_2003}  
electronic and vibrational energies can become of the same order of magnitude generating scenarios
where novel effects may emerge.\cite{Hewson_2002,Flensberg03,Braig2003,MitraAM04,Cornaglia2004,vonOppen2005,PaaskeF05,Mravlje2005,Wegewijs2005,Cornaglia2005b,Al-hassaniehBMD05,Koch2006,Balseiro2006,Mravlje2006,Galperin2006,Zazunov2006,Kikoin2006,Galperin2007}

The effect of the coupling between electronic excitations and vibronic
states in molecular transistors depends on the symmetry and frequency of the
vibrating mode and on the strength of the coupling. Symmetric modes with a
Holstein coupling between quantized vibrations and electronic levels may
strongly renormalize the molecular parameters reducing charging energies
\cite{Hewson_2002} and producing anomalous behavior of the Kondo temperature 
versus applied gate voltages.\cite{Balseiro2006} Asymmetric modes coupled 
to the tunnel-barriers parameters, like the center
of mass motion mode, can dynamically open new channels for electron
transport.\cite{Balseiro2006} 
Due to the large variety of magnetic molecules that could be
incorporated in molecular circuits with different chemical environments, it
is important to characterize and understand the behavior of molecular
transistors with different vibronic and electronic structures. 

In this work we discuss the electronic transport through a magnetic molecule with spin $S\smeq1/2$ and $S\smeq1$ in the Kondo limit. We use slave boson techniques and include the electron-phonon interaction with different coupling constants. We show that in the presence of vibrational modes, the low temperature behavior in the case of $S\smeq1$ corresponds to a two channel Kondo problem. In particular, the zero temperature conductance goes to zero.

The paper is organized as follows: in Sec. II we introduce a model Hamiltonian for the system and use the Schieffer-Wolff transformation to obtain the low energy Hamiltonian of the molecular system for both $S\smeq1/2$ and $S\smeq1$. In Sec. III we use slave boson techniques to find the ground state properties of the system in both cases and calculate the conductance. We summarize in Sec. IV.

\section{The model}

We first present the case of a spin $S\smeq1/2$ molecule described by the
Holstein-Anderson model. The Hamiltonian reads $H\smeq H_{0}+H_{\mathrm{e-ph}}+H_{\mathrm{hyb}}$,
where $H_{0}$ describes the molecular degrees of freedom and the electronic
excitations of the two leads
\begin{eqnarray}\label{eq:H0}
\nonumber
H_{0} &\smeq&\sum_{\sigma }\varepsilon _{M}d_{\sigma }^{\dagger}d_{\sigma }^{ }+\frac{U}{2}%
\sum_{\sigma ,\sigma ^{\prime }}d_{\sigma }^{\dagger}d_{\sigma }^{ }d_{\sigma ^{\prime
}}^{\dagger}d_{\sigma ^{\prime }}^{ } \\
&&+\sum_{\alpha ,k,\sigma }\varepsilon _{k}c_{\alpha k\sigma }^{\dagger}c_{\alpha
k\sigma }^{ }+\hbar \omega _{0}\,a^{\dagger}a^{ }\,,
\end{eqnarray}
where $d_{\sigma }^{\dagger}$ creates an electron in a molecular orbital with
energy $\varepsilon _{M}$, $U$ is the intramolecular Coulomb repulsion, $%
c_{\alpha k\sigma }^{\dagger}$ creates an electron in the $k$ mode of the $\alpha $
lead where $\alpha \smeq L$,$R$ stands for the left and right lead,
respectively.\ The last term in Eq.~(\ref{eq:H0}) describes the molecular vibronic mode of
frequency $\omega _{0}$. The term $H_{\mathrm{e-ph}}$ in the Hamiltonian describes the
Holstein coupling between the electrons in the molecular orbital and the
molecular vibrations,
\begin{equation}
H_{\mathrm{e-ph}}\smeq-\lambda (a^{\dagger}+a)\left(\sum_{\sigma }d_{\sigma }^{\dagger}d_{\sigma }^{}-1\right).
\end{equation}
\begin{figure}
\includegraphics[width=0.45\textwidth]{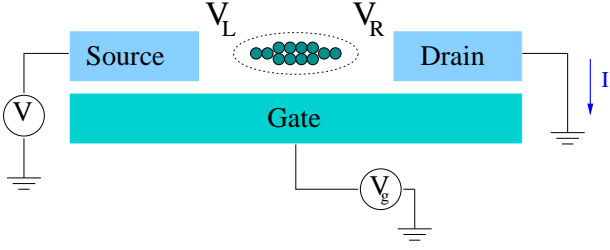}
\caption{Scheme of a molecular device.}
\end{figure}
Finally, the leads-molecule hybridization term is 
\begin{eqnarray}
H_{\mathrm{hyb}}&\smeq&\frac{1}{\sqrt{N}}\sum_{\alpha ,k,\sigma }\widehat{V}_{\alpha
}\left(d_{\sigma }^{\dagger}c_{\alpha k\sigma }+c_{\alpha k\sigma }^{\dagger}d_{\sigma }\right)\\
\nonumber
&\smeq&\frac{1}{\sqrt{N}}\sum_{\alpha ,k,\sigma }V_{\alpha }\left(1\!+\!g_{\alpha }(a^{\dagger}\!+\!a)\right)
\left(d_{\sigma }^{\dagger}c_{\alpha k\sigma }+c_{\alpha k\sigma }^{\dagger}d_{\sigma }\right)
\end{eqnarray}
where $N$ is the number of band states.
For some special relations between the parameters of the Hamiltonian, the model reduces to known and, in some cases, well studied models:

{\it i)} $V_{R}\smeq V_{L}$, $\lambda \neq 0$ and $g_{R}\smeq g_{L}\smeq0$ corresponds to
a molecule with a Holstein mode in a symmetric environment.
\cite{Hewson_2002,Cornaglia2004,Cornaglia2005b}

{\it ii)} $V_{R}\smeq V_{L}${\it , }$\lambda \neq 0$ with $g_{R}\smeq g_{L}\neq 0$ is
a generalized Anderson-Holstein model in which vibrations modify the
molecular energies and the tunneling barriers.\cite{Cornaglia2005a}

{\it iii)} $V_{R}\smeq V_{L}$, $g_{R}\smeq -g_{L}$ and $\lambda \smeq 0$ corresponds to a
molecule with inversion symmetry and a center of mass motion. \cite{Al-hassaniehBMD05,Balseiro2006}

{\it iv)} $V_{R}\smeq V_{L}$, and $g_{R}\smeq -g_{L}$ {\it \ }with $\lambda \neq 0$
describes a molecule with no inversion symmetry and a center of mass mode.
\cite{Al-hassaniehBMD05,Mravlje2006}

In the Kondo regime ($-\widetilde{U}<\widetilde{\varepsilon} _{M}<0$), when the number of electrons in the molecule is well defined, the
charge excitations can be integrated out by means of a Schrieffer--Wolff
transformation.\cite{Schrieffer1966} The resulting Hamiltonian is 
\begin{equation}
H\smeq \sum_{k,\alpha }\varepsilon _{k}\psi _{\alpha k}^{\dagger}\psi _{\alpha k}^{}+H_K 
\end{equation}
with
\begin{equation}
H_{K}\smeq \sum_{\alpha ,\beta \smeq R,L}J_{\alpha \beta }\frac{1}{N}\sum_{k,k^{\prime
}}{\bm S}\cdot\frac{1}{2}\psi _{\alpha k}^{\dagger}{\bm \sigma }\psi _{\beta
k^{^{\prime }}}.
\end{equation}
Here $\psi _{\alpha k}^{\dagger}\smeq (c_{\alpha k\uparrow }^{\dagger},c_{\alpha k\downarrow
}^{\dagger})$ is a spinor corresponding to the $\alpha $ lead, ${\bm S}$ is the
spin operator associated with the molecular orbital and ${\bm \sigma }$ are
the Pauli matrices. The coupling constants are
\begin{equation}
J_{\alpha \alpha }\smeq -2V_{\alpha }^{2}\left[ \sum_{n\smeq 0}^{\infty}\frac{\left( \gamma
_{0n}+g_{\alpha }\gamma _{1n}\right) ^{2}}{\widetilde{\varepsilon }%
_{M}-\hbar \omega _{0}n}+\sum_{n=0}^{\infty}\frac{\left( \gamma _{0n}-g_{\alpha
}\gamma _{1n}\right) ^{2}}{-\widetilde{U}-\widetilde{\varepsilon }_{M}-\hbar
\omega _{0}n}\right]
\label{Jaa} 
\end{equation}
and 
\begin{eqnarray}
\nonumber
J_{RL} &\smeq &-2V_{L}V_{R}\sum_{n=0}^{\infty}\frac{\left( \gamma _{0n}+g_{R}\gamma
_{1n}\right) \left( \gamma _{0n}+g_{L}\gamma _{1n}\right) }{\widetilde{%
\varepsilon }_{M}-\hbar \omega _{0}n} \\
&&-2V_{L}V_{R}\sum_{n=0}^{\infty}\frac{\left( \gamma _{0n}-g_{L}\gamma _{1n}\right)
\left( \gamma _{0n}-g_{R}\gamma _{1n}\right) }{-\widetilde{U}-\widetilde{%
\varepsilon }_{M}-\hbar \omega _{0}n}
\label{JRL}
\end{eqnarray}
where $\gamma _{0n}\smeq (1/\sqrt{n!})(\lambda /\hbar \omega _{0})^{n}e^{-(\lambda
/\hbar \omega _{0})^{2}/2}$, $\gamma _{1n}\smeq \gamma _{0n}(n-(\lambda /\hbar
\omega _{0})^{2})\hbar \omega _{0}/\lambda $, $%
\widetilde{\varepsilon }_{M}\smeq \varepsilon _{M}+\lambda ^{2}/\hbar \omega _{0}$%
, $\widetilde{U}\smeq U-2\lambda ^{2}/\hbar \omega _{0}$ and $J_{LR}\smeq J_{RL}$. The
electron phonon interaction renormalizes the molecular parameters $%
\varepsilon _{M}$ and $U$: the renormalized energy $\widetilde{\varepsilon }%
_{M}$ increases with  $\lambda$ while the intramolecular interaction $\widetilde{U}$ decreases. This
naturally leads to an increase of the Kondo temperature while it reduces the range of gate
voltages where the charge state with one electron is stable.\cite{Cornaglia2004} For large enough $\lambda$,
$\widetilde{U}$ becomes negative, which corresponds to an electron-phonon induced attractive interaction. In what follows we consider the case $\widetilde{U}>0$. In this work, we neglect the potential scattering term generated by the Schrieffer-Wolff transformation.

It is convenient to make a rotation in the $\left[ L,R\right] $ space to
eliminate cross terms. We then define the \textit{even} ($e$) and \textit{odd} ($o$) operators,
\begin{eqnarray}
\nonumber
c_{ek\sigma } &\smeq &u\,c_{Rk\sigma }+v\,c_{Lk\sigma } \\
c_{ok\sigma } &\smeq &v\,c_{Rk\sigma }-u\,c_{Lk\sigma }
\end{eqnarray}
with $u^{2}\smeq (1+\sqrt{1-F})/2$, $v^{2}\smeq (1-\sqrt{1-F})/2$ and $%
F\smeq 4J_{RL}^{2}/\left[ \left( J_{RR}-J_{LL}\right) ^{2}+4J_{RL}^{2}\right] $.
In terms of these operators the Hamiltonian reads
\begin{equation}
H_{K}\smeq \sum_{\eta =e,o}J_{\eta }\frac{1}{N}\sum_{k,k^{\prime }}{\bm S}\cdot\frac{1%
}{2}\psi _{\eta k}^{\dagger}{\bm \sigma }\psi _{\eta k^{^{\prime }}}\,,
\label{TCK}
\end{equation}
where 
\begin{equation}
J_{\eta }\smeq \frac{1}{2}\left(J_{RR}+J_{LL}\pm\sqrt{\left(
J_{RR}-J_{LL}\right) ^{2}+4J_{RL}^{2}}\right)
\label{Jeta} 
\end{equation}
with the $+$ and $-$ signs corresponding to the $e$-channel and $o$-channel,
respectively. The Hamiltonian (\ref{TCK}) corresponds to a two channel Kondo model.
In our case the two channels character is a consequence of the dynamical nature of the hybridization operator
and the lack of right and left symmetry: if $g_{R}\smeq g_{L}$ then $%
J_{RL}^{2}\smeq J_{RR}J_{LL}$ and $J_{o}\smeq 0$ . For $g_{R}\neq g_{L}$ the system
has always two channels coupled to the molecule's spin. However, as in general $%
J_{e}>J_{o}$, the resulting low temperature behavior is dominated by the $e$-channel.
In the next section we discuss this behavior in some detail.

More interesting is the case of a $S\smeq1$ molecule. Such a case can be
described by including an additional spin $\ {\bm S}_{0}\smeq1/2$ that
couples ferromagnetically to the spin of the electrons in the hybridized
molecular orbital described by the first terms in Eq. (1). Then, the
Hamiltonian includes a term $-J_{H}$ ${\bm S}_{0}\cdot{\bm S}$. The total spin
of the molecule is either zero or one and for large values of the Hund-like
coupling $J_{H}$ the singlet state can be neglected. Projecting the system
onto the total spin $S\smeq1$ state, the Schrieffer--Wolff transformation gives
again the Hamiltonian of Eq.(\ref{TCK}), where now the molecular spin operator
corresponds to $S\smeq1$ and the coupling constants are half the
expressions obtained in the previous case with $\varepsilon _{M}$
replaced with $\varepsilon _{M}-3J_{H}/4$. 

In the context of semiconducting quantum dots, it has been shown that the zero temperature conductance for a $S\smeq1$ dot coupled with two channels is $G\smeq G_{0}\sin ^{2}(\delta _{1}\smmi\delta _{2})$
were $\delta _{1}$ and $\delta _{2}$ are the scattering phase shifts at the
Fermi level for the two channels.\cite{Pustilnik2001}
Then, we expect molecules with $S\smeq1/2$ and $S\smeq1$ to behave quite differently if phonons open the $o$-channel. Below we present a slave boson mean-field approach to
describe this situation. 

Before finishing this section we would like to emphasize that in our model there is only one hybridized orbital so that the two channels character arises only from the coupling to the phonons. This is different from the case described in Ref. [\onlinecite{Pustilnik2001}] where the authors considered two different orbitals coupled to the leads. Our model then, corresponds to the simplest scenario that allows us to discuss the effect of the electron-phonon coupling. In general, when more than one molecular orbital is coupled to the leads, there would be more than two channels involved, which presumably will lead to a more complex behavior.

\section{Slave boson mean field theory}

For the sake of completeness, we start with a brief discussion of the spin
$S\smeq1/2$ case. We review the slave boson mean field theory and calculate the
conductance. Although the results for this case are well known, they are presented as a guide to the more interesting case of a spin $S\smeq1$ molecule.

\subsection{The $S=1/2$ case}

Following standard procedures, we use a fermionic representation for the
molecular spin,
\begin{equation}
S_{z}\smeq \frac{1}{2}(f_{\uparrow }^{\dagger}f_{\uparrow }-f_{\downarrow
}^{\dagger}f_{\downarrow })\text{, }\qquad S^{+}\smeq f_{\uparrow }^{\dagger}f_{\downarrow }\text{,
}\qquad S^{-}\smeq f_{\downarrow }^{\dagger}f_{\uparrow }\,.
\label{defS}
\end{equation}
In terms of these operator, the Kondo Hamiltonian reads 
\begin{eqnarray}
\nonumber
H_{K}\!&\!\smeq \!&\!\frac{1}{N}\!\sum_{\eta =e,o}\!J_{\eta }\sum_{k,k^{\prime }}\frac{1}{4}%
(f_{\uparrow }^{\dagger}f_{\uparrow }\!-\!f_{\downarrow }^{\dagger}f_{\downarrow })(c_{\eta
k\uparrow }^{\dagger}c_{\eta k^{\prime }\uparrow }\!-\!c_{\eta k\downarrow
}^{\dagger}c_{\eta k^{\prime }\downarrow }) \\
&&+\frac{1}{2}(f_{\uparrow }^{\dagger}f_{\downarrow }c_{\eta k\downarrow
}^{\dagger}c_{\eta k^{\prime }\uparrow }+f_{\downarrow }^{\dagger}f_{\uparrow }c_{\eta
k\uparrow }^{\dagger}c_{\eta k^{\prime }\downarrow })\,.
\end{eqnarray}
The mean field approach corresponds to define bosonic fields $\lambda _{\eta
\sigma }\smeq 1/\sqrt{N}\sum_{k}c_{\eta k\sigma }^{\dagger}f_{\sigma }$ and evaluate
the physical quantities at the mean field level in these fields. The resulting mean field
Hamiltonian can be put as\cite{PhysRevB.56.11820} 
\begin{equation}
H_{K}\smeq -\sum_{\eta =e,o}\frac{J_{\eta }}{\sqrt{N}}\sum_{k,\sigma }
\left\langle
\lambda _{\eta -\sigma }\right\rangle f_{\sigma }^{\dagger}c_{\eta k\sigma
}+H.c. 
\label{HK1C}
\end{equation}
where $\left\langle ..\right\rangle $ indicates the thermal average. The
solution must satisfy the condition $\left\langle f_{\uparrow
}^{\dagger}f_{\uparrow }+f_{\downarrow }^{\dagger}f_{\downarrow }\right\rangle \smeq 1$ that
is imposed by introducing a Lagrange multiplier. The procedure is well
justified only when the original model is extended for the spin variable $%
\sigma $ to take ${\cal N}$ different values with ${\cal N}\rightarrow
\infty $. However, it is well known that the approximation captures
essentially the correct low temperature physics even for ${\mathcal N}\smeq 2$. 
In the absence of an external magnetic field we may take $\left\langle
\lambda _{\eta \uparrow }\right\rangle \smeq \left\langle \lambda _{\eta
\downarrow }\right\rangle \equiv \Lambda _{\eta }$. The self-consistent
solutions for the bosonic fields are 
\begin{equation}
\Lambda _{\eta }\smeq -J_{\eta }\Lambda _{\eta }\rho
\int_{-D}^{D}d\omega f(\omega )\frac{\omega }{\omega ^{2}+\Gamma ^{2}}\,,
\label{lambda} 
\end{equation}
where $f(\omega )$ is the Fermi function, $\rho \smeq 1/2D$ is the leads density
of states per spin and 
\begin{equation}
\Gamma \equiv \Gamma _{e}+\Gamma _{o}\smeq \pi \rho \lbrack (J_{e}\Lambda
_{e})^{2}+(J_{o}\Lambda _{o})^{2}]\,.
\end{equation}
For $J_{e}\neq J_{o}$ these equations do not have a solution with non-zero values for both $%
\Lambda _{e}$ and $\Lambda _{o}$. At zero temperature we have 
\begin{equation}
\Lambda _{\eta } \smeq \sqrt{\frac{2}{\pi }}\frac{D}{J_{\eta }}e^{-\frac{D}{%
J_{\eta }}}\,, \qquad \Lambda _{-\eta } \smeq 0
\end{equation}
where $\eta $ ($-\eta $) corresponds to the largest (smaller) Kondo coupling 
$J_{\eta }.$ This solution describes the situation in which one channel
completely decouples from the molecular spin. In the renormalization group
language the ratio between the largest and smaller couplings flows to
infinity and the Kondo screening is due only to the channel with the largest
coupling. Note that for the particular case $J_{e}\smeq J_{o}$, the overscreening
of the $S\smeq1/2$ spin requires a more elaborate approximation.\cite{Hewson-book}

In what follows we assume that $J_{e}>J_{o}$. The temperature dependence of
the Kondo correlation energy, defined as $E_{K}\smeq \left\langle
H_{K}\right\rangle \smeq -4J_{e}\Lambda _{e}^{2}$, is shown in figure 2. At a
temperature $T_{Ke}$ the correlation energy goes to zero in a singular way.
This is known to be an artifact of the approximation, $T_{Ke}$ should be
interpreted as a crossover temperature between a high temperature regime
where the spin is essential unscreened and a low temperature Fermi liquid
regime. According to Eq. (\ref{lambda}) the Kondo temperature is given by the solution of
the following equation
\begin{equation}
1\smeq \frac{J_{e}}{2}\rho \int_{-D}^{D}d\omega \frac{1}{\omega }\tanh (\frac{%
\omega }{2T_{Ke}}).
\end{equation}
The linear conductance through the molecule is calculated using a Landauer
approach. At the mean field level described by Hamiltonian (\ref{HK1C}), the problem
reduces to a single resonant level centered at the Fermi energy.
This structure represents the Kondo resonance that leads to resonant
tunneling between the source and the drain contacts.

\begin{figure}
\includegraphics[width=.45\textwidth]{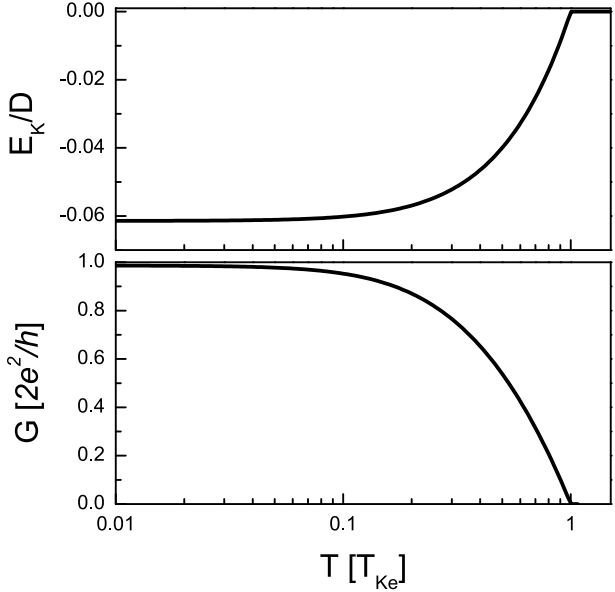}
\caption{Temperature dependence of the correlation energy (top) and the conductance (bottom) in
the $S\smeq 1/2$ case. Parameters: $D\smeq 2$, $\rho J\smeq 0.22$ and $V_L\smeq V_R$}
\end{figure}

The transmission function is given by $T(\omega )\smeq \Gamma _{R}G^{r}\Gamma _{L}G^{a}$
with $\Gamma _{R}\smeq u^{2}\Gamma _{e}$ and $\Gamma _{L}\smeq v^{2}\Gamma _{e}$, $%
G^{r}$ and $G^{a}$ are the retarded and the advanced propagators of the $f$-fermions,
respectively. In terms of $T(\omega)$ the conductance is then,
\begin{eqnarray}
\nonumber
G &\smeq &\frac{2e^{2}}{h}\int d\omega \left( -\frac{\partial f(\omega )}{%
\partial \omega }\right) T(\omega ) \\
&\smeq &\frac{2e^{2}}{h}\int d\omega \left( -\frac{\partial f(\omega )}{\partial
\omega }\right) 4u^{2}v^{2}\frac{\Gamma _{e}^{2}}{\omega ^{2}+\Gamma _{e}^{2}%
}\,.
\end{eqnarray}
The last expression was obtained by using the Green functions corresponding to a resonant level of width $\Gamma_e$.
For a system with inversion symmetry, we have $V_{L}\smeq V_{R}$ and $4u^{2}v^{2}\smeq 1$.
In this situation the zero temperature conductance is the quantum unit of
conductance $2e^{2}/h$. As the temperature increases $\Gamma _{e}$ decreases
and goes to zero at the Kondo temperature $T_{Ke}$. The temperature
dependence of the conductance is shown in figure 2. As already mentioned, the
singular behavior at $T_{Ke}$ is to be interpreted as a crossover. In fact,
above the Kondo temperature the conductance is small but non-zero and can be
calculated using perturbation theory.\cite{PhysRevLett.89.206602} 

\subsection{The $S=1$ case}

From the many possible representations of a spin one operators, we use a two
spin one-half fermion representation that has proved to give good results
when treated at the mean field level. Following Refs. [\onlinecite{PhysRevB.65.054413,Dolores}] we define two states
with quantum numbers $i\smeq 1,2$ and $\sigma \smeq \uparrow ,\downarrow $ . The
operators $f_{i\sigma }^{\dagger}$ create a fermion in these states. The physical Hilbert space
corresponds to the subspace with $n_{i}\smeq n_{i\uparrow }+n_{i\downarrow }\smeq 1$, with
$n_{i\sigma }\smeq f_{i\sigma }^{\dagger}f_{i\sigma }$. The spins of the states $1$ and $2$ are coupled ferromagnetically with a exchange constant $J_H$. Within
this subspace, the three components of the spin one operator are given by 
\begin{eqnarray}
\nonumber
S_{z} &\smeq &n_{1\uparrow }n_{2\uparrow }-n_{1\downarrow }n_{2\downarrow } \\
\nonumber
S^{+} &\smeq &n_{1}S_{2}^{+}+S_{1}^{+}n_{2} \\
S^{-} &\smeq &S_{1}^{-}n_{2}+n_{1}S_{2}^{-}\,,
\end{eqnarray}
here $S_{i}^{\pm }$ are the (raising/lowering) spin operators of state $i$, which are defined as in Eq. (\ref{defS}). 

As mentioned above, the effective Hamiltonian $H_K$ for this case is described again by 
Eq. (\ref{TCK}), but the definition of the coupling constant $J_\eta$ is obtained by replacing $\varepsilon _{M}\rightarrow \varepsilon _{M}-3J_{H}/4$ in Eqs. (\ref{Jaa}) and (\ref{JRL}) and dividing Eq. (\ref{Jeta}) by $2$. 
Defining bosonic
fields as 
\begin{equation}
\lambda _{\eta \sigma }^{i}\smeq \frac{1}{\sqrt{N}}\sum_{k}c_{\eta k\sigma
}^{\dagger}f_{i\sigma }
\end{equation}
we can write the transverse part of the spin product as 
\begin{equation}
\frac{1}{N}\sum_{k,k^{\prime }}(S^{+}c_{\eta k\downarrow
}^{\dagger}c_{\eta k^{^{\prime }}\uparrow }\!+\!S^{-}c_{\eta k\uparrow }^{\dagger}c_{\eta
k^{^{\prime }}\downarrow })\!\smeq \!-\!\sum_{i\sigma }\lambda _{\eta
-\sigma }^{i}\lambda _{\eta \sigma }^{i\dagger}n_{\overline{i}}
\label{XY}
\end{equation}
with $\overline{i}\smeq (i+1,\mathop{\rm mod}2)$. As in the $S\smeq1/2$ case, we treat this term in the mean field approximation. This gives 
\begin{eqnarray}
\nonumber
-\sum_{i\sigma }\lambda _{\eta \bar{\sigma}}^{i}\lambda _{\eta \sigma
}^{i\dagger}n_{\overline{i}} &\!\simeq\!&\!-\sum_{i\sigma }(\lambda _{\eta
\bar{\sigma}}^{i}\left\langle \lambda _{\eta \sigma }^{i\dagger}\right\rangle 
\!+\!\langle \lambda _{\eta \bar{\sigma} }^{i}\rangle \lambda _{\eta
\sigma }^{i\dagger})\langle n_{\overline{i}}\rangle  \\
\nonumber
&\!+\!&\left\langle \lambda _{\eta \bar{\sigma} }^{i}\right\rangle \left\langle
\lambda _{\eta \sigma }^{i\dagger}\right\rangle n_{\overline{i}} 
-2\left\langle \lambda _{\eta \bar{\sigma}}^{i}\right\rangle \left\langle
\lambda _{\eta \sigma }^{i\dagger}\right\rangle \left\langle n_{\overline{i}%
}\right\rangle\\ 
\end{eqnarray}
where $\bar{\sigma}\smeq -\sigma$. 
\begin{figure}
\includegraphics[width=0.45\textwidth]{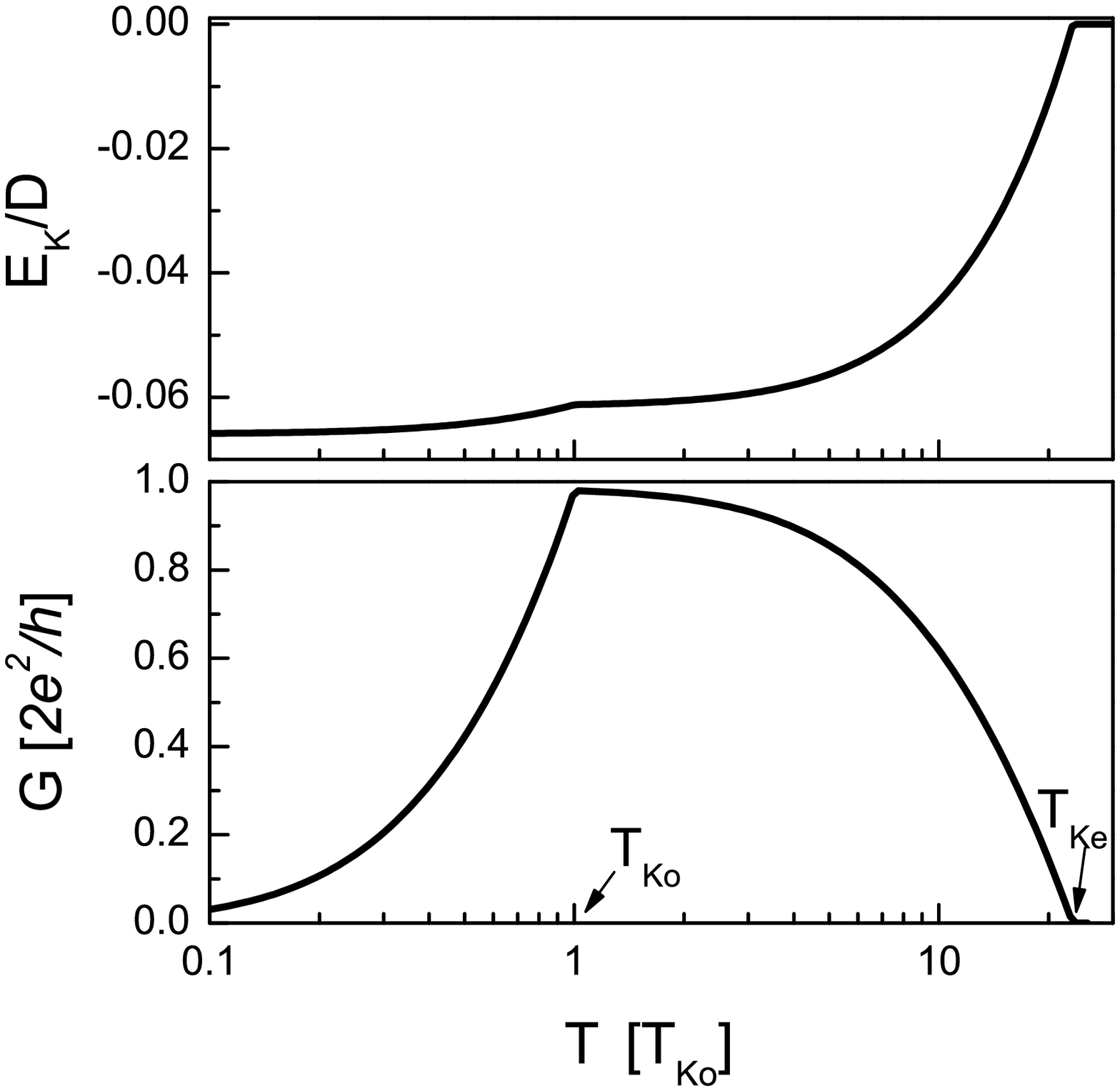}
\caption{Temperature dependence of the correlation energy (top) and the conductance (bottom) in
the $S\smeq 1$ case. Parameters: $D\smeq 2$, $\rho J_e\smeq 0.22$, $\rho J_o\smeq
0.13$}
\label{EG2C}
\end{figure}
In the absence of external magnetic fields these quantities $\left\langle
\lambda _{\eta \sigma }^{i}\right\rangle $ are spin independent. 

The
Kondo Hamiltonian is then given by 
\begin{equation}
H_{K}\smeq \sum_{\eta \smeq e,o}-\frac{J_{\eta }}{\sqrt{N}}%
\sum_{i,k,\sigma }\left\langle \lambda _{\eta \bar{\sigma}}^{i\dagger}\right\rangle
c_{\eta k\sigma }^{\dagger}f_{i\sigma }+H.c.
\label{HK-2C}
\end{equation}
where we have taken $\left\langle n_{i}\right\rangle \smeq 1$ . The total
Hamiltonian also includes a term of the form $\sum_{i}\mu _{i}(n_{i}-1)+C$ \
where $\mu _{i}$ is a Lagrange multiplier introduced to preserve the number
of $f$-fermions and $C$ is a constant. The quantities $\left\langle \lambda
_{\eta \sigma }^{i\dagger}\right\rangle $ are calculated self-consistently. The
symmetry of the Hamiltonian in the quantum number $i$ may suggest that they
are $i$-independent. However, the most general solution is 
\begin{equation}
\left\langle \lambda _{\eta \sigma }^{1+}\right\rangle \smeq \Lambda _{\eta }%
\,,\qquad\left\langle \lambda _{\eta \sigma }^{2+}\right\rangle \smeq \Lambda
_{\eta }e^{\ci\theta _{\eta }}
\end{equation}
with\ $\Lambda _{\eta }$ a real number. For the one channel case, the fase $\theta_e$
can be eliminated by a gauge transformation. However, as shown below, in
the two channel case the fase difference $\delta\theta\smeq \theta _{e}-\theta _{o}$ plays
an important role. With the notation defined above, the Hamiltonian (\ref{HK-2C}) reads
\begin{eqnarray}
\nonumber
H_{K} &\smeq &-J_{e}\Lambda _{e}\sqrt{\frac{2}{N}}\sum_{k,\sigma
}c_{ek\sigma }^{\dagger}f_{e\sigma }+H.c \\
\nonumber
&-&J_{o}\Lambda _{o}\sqrt{\frac{2}{N}}\sum_{k,\sigma }\cos\left(\frac{%
\delta \theta }{2}\right)\widetilde{c}_{ok\sigma }^{\dagger}f_{e\sigma }\\
&+&i\sin \left(\frac{\delta \theta }{2}\right)\widetilde{c}_{ok\sigma }^{\dagger}f_{o\sigma }+H.c.
\label{Hlast}
\end{eqnarray}
where $f_{e\sigma }\smeq (f_{1\sigma }+e^{\ci\theta _{e}}f_{2\sigma })/\sqrt{2}$, $%
f_{o\sigma }\!\smeq (f_{1\sigma }-e^{\ci\theta _{e}}f_{2\sigma })/\sqrt{2}$, and $%
\widetilde{c}_{ok\sigma }^{\dagger}\smeq e^{-\ci\delta \theta /2}c_{ok\sigma }^{\dagger}$.
 For $\delta \theta \smeq 0$, the two
channels are mixed with the same fermion. Then, as in the $S\smeq 1/2$ \ case, one
channel decouples and does not contribute to the energy. For $\delta \theta
\smeq \pi $ the mean field problem reduces to the case of two channels, each one
coupled with a resonant state and both contributing to lower the energy. It
can be shown that for the general case with $J_{e}\neq J_{o}$,
self-consistent solutions exist only for\ $\delta \theta \smeq 0$ or $\pm \pi $.
Without any loss of generality we can take $\theta _{e}\smeq 0$. Then, for $\delta
\theta \smeq \pm \pi $ \ we have 
\begin{equation}
H_{K}\smeq \sum_{\eta =e,o}-J_{\eta }\Lambda _{\eta }\sqrt{\frac{2}{N}}%
\sum_{k,\sigma }c_{\eta k\sigma }^{\dagger}f_{\eta \sigma }+H.c.
\end{equation}
where the self-consistent parameters $\Lambda _{\eta }$ are given by  
\begin{equation}
1\smeq -J_{\eta }\rho \int_{-D}^{D}d\omega f(\omega )\frac{\omega }{%
\omega ^{2}+\Gamma _{\eta }^{2}}
\end{equation}
with $\Gamma _{\eta }\smeq \pi \rho q_{\eta }^{2}$ and $q_{\eta }\smeq J_{\eta
}\Lambda _{\eta }$. Accordingly, the zero temperature correlation energy is 
$E_{K}\smeq -4J_{e}\Lambda _{e}^{2}-4J_{o}\Lambda _{o}^{2}$.
As the two coupling constants are different, there are
two characteristic energy scales given by the two Kondo temperatures $T_{Ke}$
and $T_{Ko}$. The correlation energy as a function of temperature is shown in figure \ref{EG2C}. 

The conductance can be easily evaluated within the mean field approximation following the same line as in the $S=1/2$ case. The transmission function in this case is given by 
$T(\omega )\smeq \mathrm{Tr}[{\bm \Gamma }_{L}{\bm G}^{a}{\bm \Gamma }_{R}{\bm G}^{r}]$, where the ${\bm \Gamma }$'s and the propagators are $2\times 2$ matrices since we now have two active channels coupled to the $S\smeq1$ spin. In the even and odd base we
have  
\begin{equation}
{\bm \Gamma }_{R}\smeq \pi \rho\left(
\begin{array}{ll}
q_eq_e & q_eq_o \\ 
q_oq_e & q_oq_o
\end{array}
\right)\,,\,
{\bm \Gamma }_{L}\smeq \pi \rho\left(
\begin{array}{ll}
q_eq_e & -q_eq_o \\ 
-q_oq_e & q_oq_o
\end{array}
\right).
\end{equation}

\begin{figure}
\includegraphics[width=0.45\textwidth]{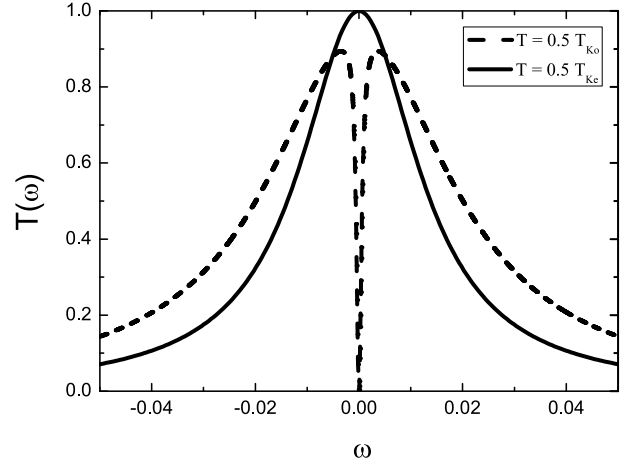}
\caption{Transmission function $T(\omega$) as a function of $\omega$ for a $S\smeq 1$ molecule. For $T_{Ko}\!<\!T\!<\!T_{Ke}$ the system shows the usual one channel Kondo resonance (solid line) while for $T\!<\!T_{Ko}$ the odd channel becomes active and a zero-transmission dip appears at $\omega\smeq 0$ (dashed line).}
\label{Aw}
\end{figure}

Finally, the conductance can be put as
\begin{equation}
G\smeq \frac{2e^{2}}{h}\int d\omega \left( -\frac{\partial f(\omega )}{\partial
\omega }\right) 4u^{2}v^{2}\left| \frac{\Gamma _{e}}{\omega +i\Gamma _{e}}-%
\frac{\Gamma _{o}}{\omega +i\Gamma _{o}}\right| ^{2}
\end{equation}
here $\Gamma _{\eta }\smeq \pi \rho q_{\eta }^{2}$ . The conductance as a function of temperature is shown in figure \ref{EG2C}. Initially the conductance increases as $T$ decreases as in the usual one-channel case but it starts decreasing towards a zero value as $T$ becomes smaller than $T_{Ko}$. This interference effect between the two channels is clearly seen as a dip at $\omega\smeq 0$ in the transmission function (Fig. \ref{Aw}). 

Interference effects in the context of `single impurity' Kondo physics have been extensively discussed in the literature.\cite{UjsaghyKSZ00,BulkaS01,CornagliaB03,Al-hassaniehBMD05,ChiappeL06} In our case the origin of the effect is clear as discussed in detail by Pustilnik and Glazman.\cite{Pustilnik2001} In the low temperature Fermi liquid regime, the mean field approach presented here allows for a simple interpretation of the interference phenomena: Hamiltonian (\ref{Hlast}) with $\delta\theta=\pi$ and non-zero $\Lambda_e$ and $\Lambda_o$ is equivalent to a two lead junction connected by two different paths. In this effective model, the phase difference of the two paths is $\pi$ leading to destructive interference and zero conductance at $T=0$.

\section{Summary}

We have analyzed the Kondo Hamiltonians for magnetic molecules with vibronic
states in molecules with spin $S\smeq 1/2$ and $S\smeq 1$ within the slave boson mean
field theory. We showed that in systems with no  $L$-$R$ symmetry in the
electron-phonon coupling, there are two channels coupled to the molecular
spin. For the case of $S\smeq 1/2$ molecules, the weakest coupled channel plays
no role; at the mean field level it is simply decoupled. For the $S\smeq 1$ case,
the two channels screen the molecular spin at different energy scales. 
To fix ideas we summarize the behavior of two simple cases with identical $V_L$ and $V_R$:

{\it a}) A Holstein-like mode with $\lambda \neq 0$ and $g_{R}\smeq g_{L}$. The
Kondo Hamiltonian includes a single channel and for the $S\smeq 1/2$ case the
slave boson mean field approximation reproduces the known results. The
electron-phonon coupling renormalizes the coupling constants. At $T_{K}$ the
conductance increases to reach the unitary limit at zero temperature. In
this case it has been shown that for large electron-phonon coupling $\lambda 
$ the Kondo temperature is weakly dependent on gate voltages due to the
anomalous dependence of coupling constants $J_{\alpha \alpha }$ with the
molecular orbital energy $\varepsilon _{M}$.\cite{Balseiro2006} For the $S\smeq 1$ case,  a single channel
can screen only half of the molecular spin. The zero temperature conductance
also reaches the quantum of conductance value $2e^{2}/h$ although strictly
speaking the system is not a Fermi liquid since there is an unscreened spin one-half at
the molecular junction.

{\it b}) A molecule with inversion symmetry and a center of mass mode, $%
\lambda \smeq 0$ and $g_{R}\smeq -g_{L}$. The resulting Kondo Hamiltonian includes two
channels with different coupling constants. For the $S\smeq 1/2$ case the usual
behavior is reproduced as one channel decouples. In this simple model
electron-phonon coupling is just due to the modulation of the tunneling
barriers or hybridizations $\widehat{V}_{\alpha }$, in the lowest order the molecular
energies and Coulomb repulsion are not renormalized. For the $S\smeq 1$ case the spin
is screened by the two active channels with two characteristic energy scales 
$T_{Ke}$ and $T_{Ko}$ . For $T_{Ko}\ll T\ll T_{Ke}$ the conductance
approaches the unitary limit and decreases for $T<T_{Ko}$ being zero at zero
temperature.
In the most general case  with $\lambda \neq 0$ , and $g_{R}\neq g_{L}$ the
conductance are reduced by the prefactor $4u^{2}v^{2}$ . 

\section{Acknowledgment}
This work was partially supported by ANPCyT Grants No 13829 and 13476 and CONICET PIP
5254. GU and PSC are members of CONICET.

\end{document}